\documentclass[journal=nalefd,manuscript=letter, layout=twocolumn]{achemso}

\usepackage{chemformula} 
\usepackage[T1]{fontenc} 

\usepackage{amssymb}
\usepackage{upgreek}
\usepackage{mathtools}
\usepackage{dsfont}
\usepackage{cases}
\usepackage{xcolor}
\usepackage{subfiles}
\usepackage{float}

\usepackage{xr}
\newcommand{\draft}{false}

\newcommand*{\degree}{$^{\circ}$}

\newcommand*{\PtC}{\ch{MeCpPt(Me)3}}
\newcommand*{\CoC}{\ch{Co2(CO)8}}

\newcommand*{\norm}[1]{\left\lVert#1\right\rVert}
\renewcommand*{\vec}[1]{\mathbf{#1}}

\author{L. Skoric}
\affiliation[Cambridge]{Cavendish Laboratory, University of Cambridge, CB3 0HE, JJ Thomson Avenue, UK}
\email{ls604@cam.ac.uk}
\author{D. Sanz-Hern\'andez}
\affiliation[Cambridge]{Cavendish Laboratory, University of Cambridge, CB3 0HE, JJ Thomson Avenue, UK}
\author{F. Meng}
\affiliation[Cambridge]{Cavendish Laboratory, University of Cambridge, CB3 0HE, JJ Thomson Avenue, UK}
\author{C. Donnelly}
\affiliation[Cambridge]{Cavendish Laboratory, University of Cambridge, CB3 0HE, JJ Thomson Avenue, UK}
\author{S. Merino-Aceituno}
\affiliation[Vienna]{Faculty of Mathematics , University of Vienna, Oskar-Morgenstern-Platz 1, 1090 Vienna, Austria}
\author{A. Fern\'andez-Pacheco}
\alsoaffiliation[Cambridge]{Cavendish Laboratory, University of Cambridge, CB3 0HE, JJ Thomson Avenue, UK}
\affiliation[Glasgow]{SUPA, School of Physics \& Astronomy, Kelvin Building,  G12 8QQ, University of Glasgow, Scotland, UK}
\email{amalio.fernandez-pacheco@glasgow.ac.uk}

\title{Layer-by-layer growth of complex-shaped three-dimensional nanostructures with focused electron beams}


\begin{document}

\begin{abstract}
The fabrication of three-dimensional (3D) nanostructures is of great interest to many areas of nanotechnology currently challenged by fundamental limitations of conventional lithography. 
One of the most promising direct-write methods for 3D nanofabrication is focused electron beam-induced deposition (FEBID), owing to its high spatial resolution and versatility. 
Here we extend FEBID to the growth of complex-shaped 3D nanostructures by combining the layer-by-layer approach of conventional macroscopic 3D printers and the proximity effect correction of electron beam lithography. This framework is based on the continuum FEBID model and is capable of adjusting for a wide range of effects present during deposition, including beam-induced heating, defocussing and gas flux anisotropies. 
We demonstrate the capabilities of our platform by fabricating free-standing nanowires, surfaces with varying curvatures and topologies, and general 3D objects, directly from standard stereolithography (STL) files and using different precursors. 
Real 3D nanoprinting as demonstrated here opens up exciting avenues for the study and exploitation of 3D nanoscale phenomena.
\end{abstract}

\section{Introduction}
The realization of three-dimensional (3D) nanoscale systems is expected to play a key role for the future progress in many areas of nanotechnology such as biology\cite{qianMicroNanofabricationTechnologies2010}, nanomagnetism\cite{fernandez-pachecoThreedimensionalNanomagnetism2017}, metamaterials\cite{kadic3DMetamaterials2019}, and micro-electro-mechanical systems (MEMS)\cite{vavassoriRemoteMagnetomechanicalNanoactuation2016}. However, conventional  lithography methods that excel in planar patterning are not well suited for complex 3D nanofabrication. The development of advanced direct-write micro- and nano-scale fabrication techniques is a possible solution to this problem currently under intense investigation. These include two-photon lithography, \cite{maruoThreedimensionalMicrofabricationTwophotonabsorbed1997,seniutinas100NmResolution2018} direct ink writing, \cite{lewisDirectInkWriting2006} laser assisted methods, \cite{lewis3DNanoprintingLaserassisted2017, takaiThreedimensionalMicrofabricationUsing2014} local dispensing of ions in liquid, \cite{hirtTemplateFree3DMicroprinting2016,momotenkoWriteRead3D2016} and focused electron and ion beam-induced deposition (FEBID and FIBID). \cite{winkler3DNanoprintingFocused2019,matsuiFocusedIonBeamChemicalVaporDepositionFIBCVD2012}

The rapid prototyping of functional geometries with sub-micrometer features is particularly important for applications in areas such as magnetism, superconductivity and metamaterials which are ruled by characteristic length scales typically in the ten to hundred nanometer range. \cite{fernandez-pachecoThreedimensionalNanomagnetism2017,espositoNanoscale3DChiral2015,bezryadinQuantumSuppressionSuperconductivity2000,belkinFormationQuantumPhase2015,zhuScalableMultiphotonCoincidence2018}.
When it comes to fabricating these geometries, FEBID has demonstrated a number of advantages\cite{hirtAdditiveManufacturingMetal2017,jesseDirectingMatterAtomicScale2016}, including resolution of a few tens of nanometers, and vertical growth rates in hundreds of nanometers per second \cite{winklerHighFidelity3DNanoprintingFocused2018}. A large number of available precursors allows for the deposition of metallic, organic, semiconducting, magnetic and superconducting materials. \cite{utkeGasassistedFocusedElectron2008,huthFocusedElectronBeam2018}. Although traditionally FEBID-fabricated materials contain a significant proportion of organic impurities, deposition of highly-pure materials has been demonstrated via different strategies, including synthesis of new precursors\cite{porratiDirectWritingCoFe2015,makiseMicrostructuralAnalysisTransport2014,winholdSuperconductivityMetallicBehavior2014}, optimization of growth conditions, \cite{pablo-navarroTuningShapeComposition2017,bernauTunableNanosynthesisComposite2010,keLowdosePatterningPlatinum2013} introduction of reactive gases during growth,\cite{shawravHighlyConductivePure2016} and post-deposition purification. \cite{serrano-ramonImprovementDomainWall2013,botmanPurificationPlatinumGold2006,botmanCreatingPureNanostructures2009}. 

These advances have already led to a number of important fundamental studies in plasmonics\cite{winklerDirectWrite3DNanoprinting2017,espositoNanoscale3DChiral2015}, photonic crystals \cite{koopsTwodimensionalPhotonicCrystals2001}, and magnetic nanowires and lattices\cite{sanz-hernandezFabricationDetectionOperation2017,sanz-hernandezFabricationScaffoldBased3D2018, fernandez-pachecoMagnetotransportPropertiesHighquality2009,wartelleTransmissionXMCDPEEMImaging2017, kellerDirectwriteFreeformBuilding2018}.
Beyond nano-prototyping, a substantial increase in throughput for this technique could be accomplished by orders-of-magnitude increases in deposition rates through optimizations of gas injection systems (GIS) \cite{friedliOptimizedMoleculeSupply2009} and deposition at cryogenic temperatures \cite{bresinDirectwrite3DNanolithography2013}, and even parallelizing via multiple beams in next-generation tools\cite{riedeselFirstDemonstration331beam2019}.

The application of FEBID for 3D nanofabrication has progressed significantly in recent years by evolving from a trial-and-error approach to systematic generation of electron beam instructions via CAD software solutions \cite{fowlkesHighFidelity3DNanoprintingFocused2018, kellerPatternGenerationDirectwrite2018}. These works make an important step towards controllable nanoscale 3D printing of networks of nanowires, correcting effectively for various phenomena present in FEBID when growing pseudo-1D elements using different  approaches\cite{winkler3DNanoprintingFocused2019}. To exploit the full potential of FEBID as a 3D fabrication tool, however, the nanoprinting protocol needs to be redefined in order to generalize it to more complex geometries. 

In this article, we have developed a general layer-by-layer framework for the additive manufacturing of 3D nanoscale objects by focused electron beams based on the FEBID continuum model, offering a new level of control over the fabrication of three-dimensional nanosystems. We present an algorithm capable of creating beam scanning patterns adjusted for proximity effects directly from conventional 3D printing stereolithography (STL) file formats. We further show that the model can take account of a range of phenomena, including beam-induced heating, defocussing and gas flux anisotropy. We demonstrate the effectiveness of the model by fabricating geometrically complex 3D structures with high fidelity and with different precursors. This work represents a significant step forward in the capabilities for 3D fabrication at the nanoscale, as well as in the simplification of the nanostructure-design process.

\section{Theoretical description of growth model}

\begin{figure*}[ht]
\centering
\includegraphics[width = \linewidth, draft=\draft]{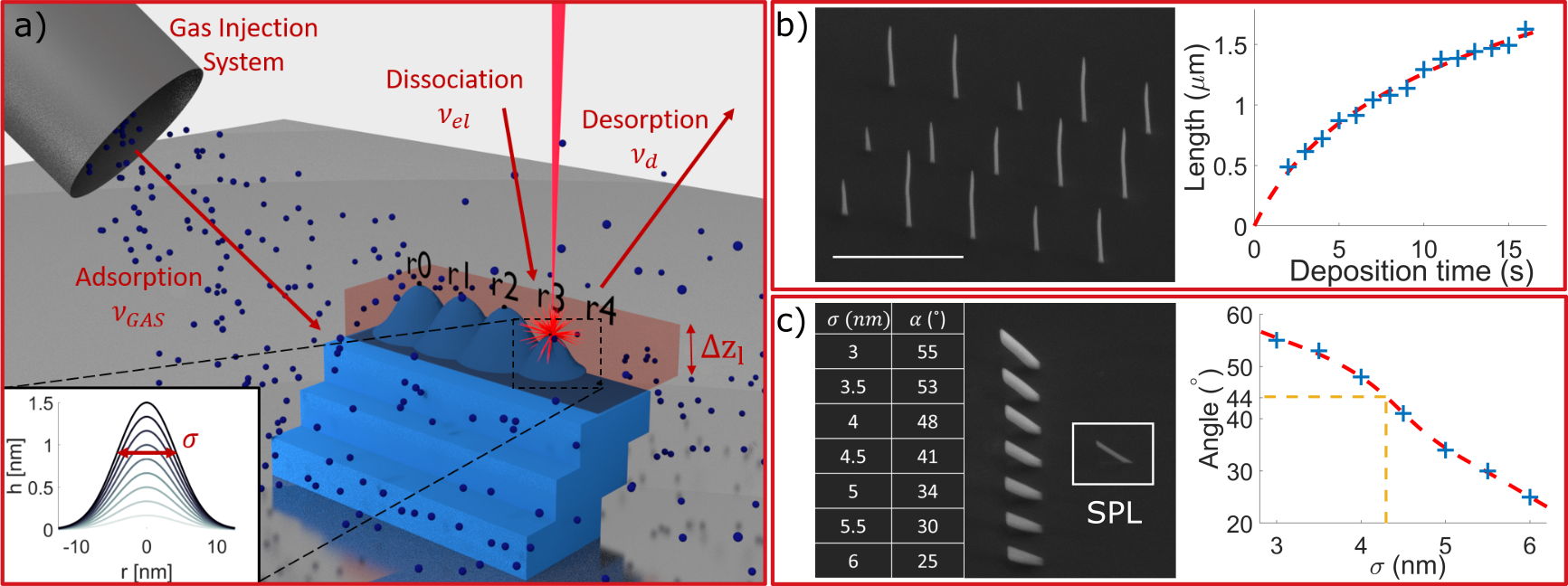}
    \caption{Schematic of FEBID 3D growth and calibration experiments. (a) Gas injection system creates a local gas coverage at the surface determined by the frequencies of adsorption ($\nu_{GAS}$), desorption ($\nu_d$) and dissociation ($\nu_{el}$). The growth in the out-of-plane direction is done in layers parallel to the focal plane of the electron beam. At each scanning point $r_i$ a deposit is created. Nearby deposits interact to form a layer of height $\Delta z_l$. Inset shows the cross section of the height evolution for a Gaussian deposit. (b) Vertical calibration used to determine growth rate and the temperature scaling factor. Here we measure $GR_0 = 300$nms$^{-1}$ and $k = 1.2$nm$^{-1}$. For fitting details, see Supporting Information 4.3. (c) The effective standard deviation of the deposit $\sigma$ is determined from short wide structures by varying $\sigma$ in the model and creating several structures. The real value of $\sigma$ is obtained by comparing these structures to a single pixel line (SPL, see inset). The plotted values are interpolated with a smoothing spline to acquire the real value of $\sigma$. Here we get $\sigma = 4.2$nm. Scale bars are $2\upmu$m. }
\label{fig:growth}
\end{figure*}

FEBID deposition is inherently a stochastic process where each electron beam dwell induces the deposition of a small number of precursor molecules. However, when averaged over the characteristic deposition times (usually in the order of a few seconds to tens of minutes), a good approximation can be achieved by describing the height $h$ of the evolving deposit as a continuous function of dwell time $t$ and radial distance $r$ from the beam center $h(t, r)$\cite{tothContinuumModelsFocused2015}. Overlapping a large number of such deposits produces out of plane growth. 

We develop a framework where, similar to conventional 3D printing, structures are grown via the deposition of thin slices/layers. In each layer, a series of deposits $i$ is made at beam scanning positions $\vec{r_i}$ with dwell times $t_i$, creating deposits $h_i(t_i, \norm{\vec{r}-\vec{r_i}})$ (see Figure \ref{fig:growth}a). 

Provided that the lateral separation between layers is low and that the individual deposits are small and closely spaced, the total height increment at any $\vec{r_i}$ is the sum over all neighboring deposits. For consistent growth, this needs to be equal to the height of a layer $l$, $\Delta z_l$, resulting in a set of equations for each layer depending on the inter-point distances and dwell times:
\begin{equation}
	\sum_{j\in N_i} h_j(t_j, \norm{\vec{r_j} - \vec{r_i}}) = \Delta z_l,
  \label{eqn:consistency}
\end{equation}
where $N_i$ is the set of all deposits of a layer $l$ in the proximity of $\vec{r_i}$ (referred to as neighboring points), and $t_j$ is the dwell time associated to each point of the layer.

In general, these equations form a non-linear optimization problem which is computationally expensive to solve. However, under the assumption that the deposit evolution can be well approximated by a separable function of time and space ${h(t, r) = s(r)f(t)}$, we can reduce Equation \ref{eqn:consistency} to a matrix equation that is numerically efficiently solvable:
\begin{equation}
  \Delta z_l = s_{ij}f(t_i),
  \label{eqn:matrixdz}
\end{equation}
where $s_{ij} = s(\norm{\vec{r_j}-\vec{r_i}})$ is a symmetric matrix, here referred to as the intra-layer proximity matrix. The dwell time values needed for consistent growth can then be calculated numerically by solving this matrix equation and inverting $f$. This framework allows us to adjust for inter-deposit proximity effects within a layer of a 3D object, similarly to ``self-consistent'' dose correction algorithms employed in electron beam lithography \cite{owenMethodsProximityEffect1990,parikhCorrectionsProximityEffects1979}. Here we treat each slice of a 3D object separately, reducing the hard problem of a general 3D growth to a set of 2D consistency equations, significantly reducing the calculation complexity. 

To use the above framework for the realistic modeling of a FEBID deposition, an accurate expression for $h(t, r)$ is needed.
In the following, we determine this function for the Langmuir FEBID model under equilibrium conditions. \cite{tothContinuumModelsFocused2015} Using the compact notation based on the characteristic FEBID frequencies that we previously developed \cite{sanz-hernandezModellingFocusedElectron2017}, the dynamics of fractional surface molecule coverage under no diffusion  $\theta$ can be described as a competition between the characteristic frequencies of gas adsorption ($\nu_{GAS}$), thermal desorption ($\nu_{d}$) and electron-induced precursor dissociation ($\nu_{el}$):
\begin{equation}
\label{eqn:thetadiff}
\frac{\partial \theta}{\partial t} = \nu_{GAS}(1-\theta) - \nu_d\theta - \nu_{el} \theta.
\end{equation} 
The corresponding growth rate ($GR$) is proportional to the amount of gas present and the electron dissociation frequency:
\begin{equation}
  \label{eqn:growthrate}
  GR = \frac{\partial h(t, r)}{\partial t} \propto \theta\nu_{el}.
\end{equation}
Equations \ref{eqn:thetadiff} and \ref{eqn:growthrate} can be solved analytically, describing the depletion of gas coverage under electron beam dissociation, and subsequent growth of the induced deposit (see Supporting Information S1.1)\cite{tothContinuumModelsFocused2015}. 

To specify the the spatial dependence of deposits, the profile of the effective flux of electrons inducing decomposition as a function of the radial distance from the electron beam $\nu_{el} = \nu_{el}(r)$ is required. For focused beams, this can be approximated by a Gaussian function\cite{utkeGasassistedFocusedElectron2008}:
\begin{equation}
  \nu_{el} = \nu_{el}^0 \exp\bigg(\frac{-r^2}{2\sigma^2}\bigg).
\end{equation} 
where $\sigma$ is the effective standard deviation, and $\nu_{el}^0$ the dissociation frequency in the center of the beam.

In general, the surface coverage will evolve during the beam dwell time, converging to an equilibrium distribution on the gas dynamics time scale (see Supporting Information S1.1):
\begin{equation}
   \tau_{dyn} = \frac{1}{\nu_{GAS}+\nu_d}.
\end{equation}

However, when growing either under electron limited conditions where the precursor depletion is small ($\nu_{el} \ll \nu_{GAS}+\nu_d$), or in the case when  the characteristic dwell time (usually on the order of 1ms) is much longer than the time-scale for convergence to equilibrium ($t_{dwell} \gg \tau_{dyn}$), 
we can assume that the system is at equilibrium at all times during the growth. This simplifies the expression for $h(t, r)$ to:
\begin{equation}
  h \propto \underbrace{\frac{\nu_{GAS}\nu_{el}(r)}{\nu_{GAS} + \nu_d + \nu_{el}(r)}}_{s(r)}t.
  \label{eqn:simplecontinuum}
\end{equation}
In this form, $h$ is written as a separable function with no further approximations. We follow this approach in our experiments as discussed below.

To allow for an analytical solution, in the above analysis we did not consider the effect of diffusion, which may become important in FEBID under mass-transport limited conditions. \cite{tothContinuumModelsFocused2015} However, in the context of Langmuir FEBID, the addition of diffusion would only alter the relevant timescales by speeding up the gas dynamics, as shown in Supporting Information S2. The separable form of deposit evolution could also be recovered in this case, albeit with a more complex form for the spatial part s(r).

\section{Experimental determination of growth parameters}

For the effective implementation of the model described above, a good approximation for $s(r)$ needs to be determined. In the following, we focus on the standard Pt-based precursor. A generalization to other precursors is discussed later.

To apply our model experimentally, key deposition parameters need to be determined for the given experimental conditions. Although in recent years a significant effort has been made towards determining these both through experiment \cite{vandorpElectronInducedDissociation2009, wnukElectronInducedSurface2009,yangProbingMorphologyEvolving2017,fowlkesFundamentalElectronPrecursorSolidInteractions2010,utkeGasassistedFocusedElectron2008,winklerHighFidelity3DNanoprintingFocused2018} and simulations \cite{smithNanoscaleThreedimensionalMonte2007,fowlkesSimulationGuided3DNanomanufacturing2016,mutungaImpactElectronBeamHeating2019,sushkoMolecularDynamicsIrradiation2016,shenPhysisorptionOrganometallicPlatinum2012}, the size of the parameter space still prevents confident ab-initio growth predictions. Here we demonstrate a set of experiments (Figures \ref{fig:growth}b and \ref{fig:growth}c) designed to find the minimal set of parameters which represent our growth conditions, enabling a simple calibration and preventing overfitting. 

We measure the growth parameters by building a set of standardized structures, isolating the dependence of the growth rate to particular factors.
Firstly, to determine growth rate at the center of the beam, a set of spot depositions is built in parallel with different total deposition times (Figure \ref{fig:growth}b). The resulting lengths of the vertical nanowires are measured and plotted against the deposition times.

Under the experimental conditions used in this paper (see Experimental), we have observed no more than 7\% variation in the average growth rate when varying the dwell and refresh times used in 3D FEBID deposition (see Supporting Information S3). Therefore, we can safely assume that Equation \ref{eqn:simplecontinuum} provides a good approximation of the deposition.

The sublinear dependence of nanowire length on the deposition times can thus be understood via the enhancement of thermal desorption due to beam-induced heating $\nu_d$ \cite{mutungaImpactElectronBeamHeating2019}.  The absolute changes in local temperature during deposition are not expected to be large compared to room temperature, being usually in the order of 10K \cite{mutungaImpactElectronBeamHeating2019,bouscaudEstimationElectronBeaminduced2012}. We can thus expand the desorption temperature dependence to first order:
\begin{equation}
  \nu_d = \nu_0 \exp\bigg(-\frac{E_d}{k_B(T_0+T_{el})}\bigg) \approx \nu_d^0 e^{\beta T_{el}},
  \label{eqn:des_approx}
\end{equation}
where $\nu_0$ is the desorption attempt frequency, $T_0$ is the base deposition temperature, $T_{el} \ll T_0$ is the change in temperature due to electron beam heating, $\nu_d^0$ is the desorption at $T_0$, and $\beta = E_d/(k_BT_0^2)$.

The significant effect of local temperature, together with low equilibrium precursor coverage in Pt FEBID \cite{friedliMassSensorSitu2007}, implies a strong dependence of the growth rates on desorption.  Therefore, even though in general we allow for the full form of the equilibrium growth rate (Equation \ref{eqn:simplecontinuum}) in our model, we have found that the desorption dominated regime ($\nu_d \gg \nu_{GAS}+\nu_{el}$)\cite{sanz-hernandezModellingFocusedElectron2017} results in the best fit to our experimental conditions. 

Under the assumption that the incident beam heating rate is the same at all dwell points and that the characteristic thermal transport time scale is expected to be short relative to the dwell times\cite{mutungaImpactElectronBeamHeating2019}, a  quasi-steady-state temperature can be assumed for every dwell. Hence, the temperature variation during deposition is fully captured by a geometrical factor $R_T$, giving account for the resistance of the 3D structure to heat transport at the point of scanning ($T_{el}\propto R_T$)\cite{mutungaImpactElectronBeamHeating2019}.

Considering the effect of beam heating in the desorption-dominated regime, Equation \ref{eqn:simplecontinuum} gives a Gaussian deposit (see inset in Figure \ref{fig:growth}a):
\begin{equation}
\label{eqn:Ptdep}
  h(t, r) = GR_0 e^{-kR_T}\exp\bigg(\frac{-r^2}{2\sigma^2}\bigg)t,
\end{equation}
where the growth rate in the center of the beam at a base deposition temperature $GR_0$, thermal resistance scaling factor $k$, and the standard deviation of the deposit $\sigma$ are parameters that need to be determined. The exponential temperature dependence for desorption (Equation \ref{eqn:des_approx}) is expressed here through the geometry-dependent factor $e^{-kR_T}$. For details about the implementation of the resistance model and the derivation of Equation \ref{eqn:Ptdep} can be found in Supporting Information S4.2.

This experiment, involving vertical deposits, allows us to remove the potential influence of the geometry of the structure and proximity effects on the resulting growth. We extract the expected dependence of lengths $L(t)$ on deposition times $t$ based on Equation \ref{eqn:Ptdep}, with  $w_0$ as their width, and growth rate $GR_0$ and resistance scaling factor $k$  as fitting parameters (see Supporting Information S4.3 for the full derivation):
\begin{equation}
  L(t) = \frac{w_0}{k}\log\bigg(\frac{kGR_0t}{w_0}+1\bigg).
\end{equation}

A second experiment (Figure \ref{fig:growth}c) is designed to obtain a value for $\sigma$. For this, a set of nominally 150nm wide straight nanowires tilted at 45\degree\ is built using the algorithm, while varying the value of $\sigma$ in the model with the goal of matching the angle of a single pixel thin nanowire built with the same conditions. For the Pt precursor, we find $GR_0 \approx 100$-$300$nms$^{-1}$, and $\sigma \approx 4$-$5$nm, depending on the SEM system and the conditions inside the microscope chamber, such as temperature, base pressure, and precursor flux.

\begin{figure}[ht]
\centering
\includegraphics[width = \linewidth, draft=\draft]{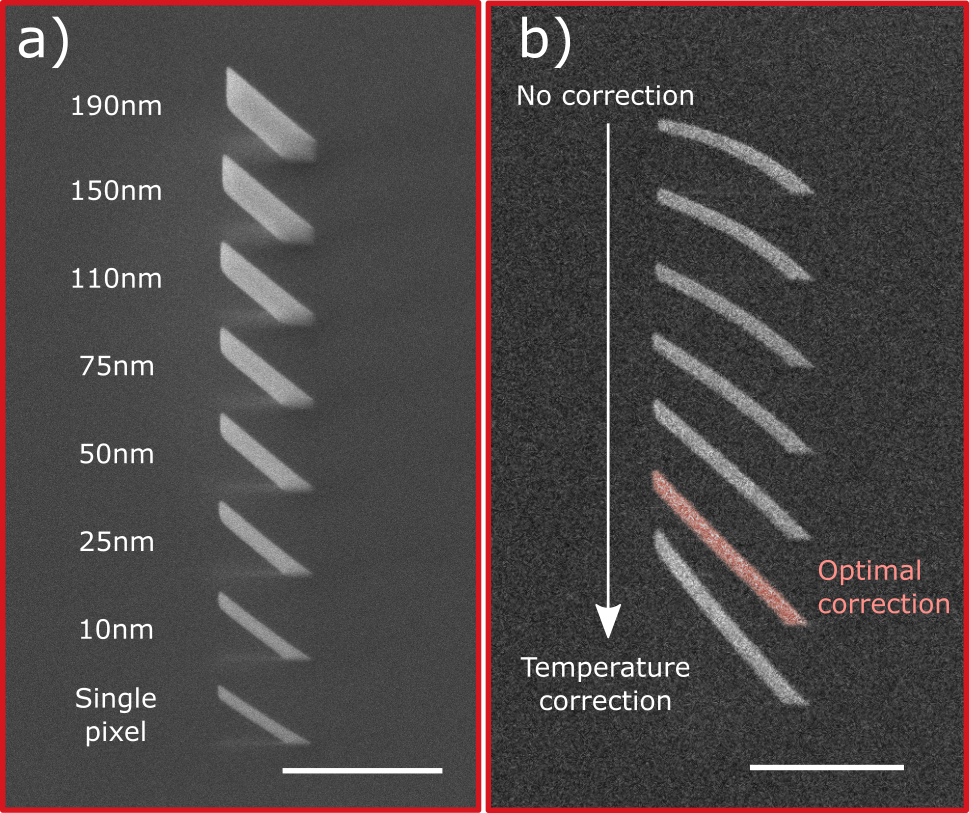}
\caption{(a) A series of nanowires and their nominal widths, demonstrating that the algorithm accounts well for proximity effects. (b) Nanowires built with increasing temperature correction factor $k$, showing how straight growing nanowires can be recovered with optimal correction acquired from calibration colored in red. Both images are side-views taken at 45\degree \ tilt. Scale bars are 1$\upmu$m.}
\label{fig:testing}
\end{figure}

This two-step calibration procedure allows us to uniquely determine all growth parameters required for implementing an algorithm to pattern 3D complex structures. We further tested the validity of the calibration protocol by growing two sets of additional test structures: nanowires with constant out-of-plane angle and variable controllable width (Figure \ref{fig:testing}a), and nominally-straight nanowires, where the $k$ value obtained by the fit in Figure \ref{fig:growth}b leads to the growth of a straight structure (see "Optimal correction" in Figure \ref{fig:testing}b).

Apart from temperature, we studied other potentially-relevant effects such as electron beam defocussing and gas flux anisotropy. These have been found to have a relatively small effect for our experimental conditions, but can be important for very fine growth in certain structures and deposition regimes. If required, further corrections of this type can be easily implemented in our model at the expense of introducing additional parameters that require independent calibration (see Supporting Information S5).

\section{3D printing algorithm}

\begin{figure*}[ht]
\centering
\includegraphics[width = 0.75\linewidth, draft=\draft]{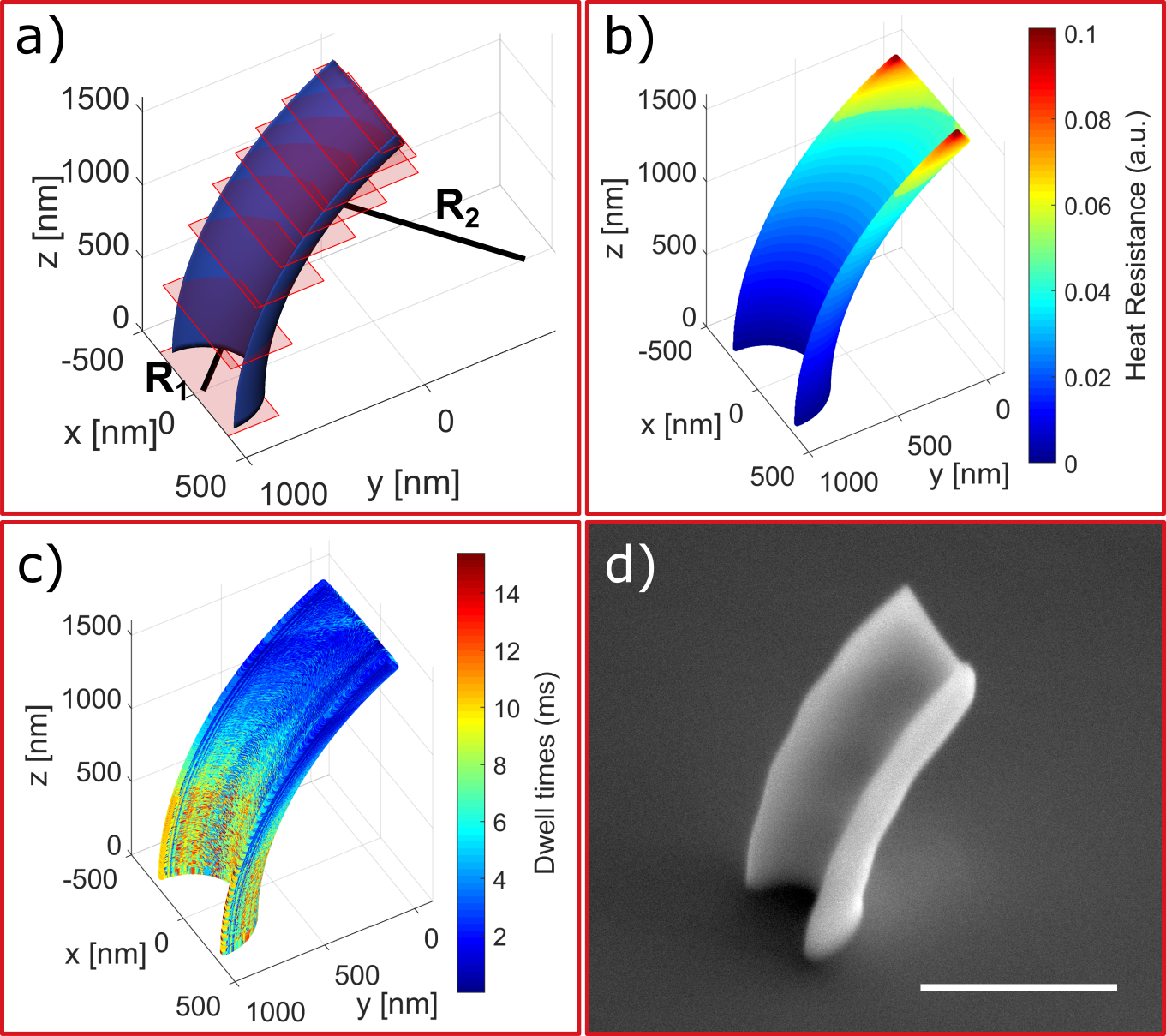}
\caption{Workflow of the algorithm to create 3D complex objects by FEBID. (a) Design of a curved concave plane with two curvature radii ($R_1 = 300$nm, $R_2 = 1500$nm). The structure is sliced in layers and a dense mesh of regularly spaced points is generated for each layer. For clarity, every 75th slice is displayed. 
(b) Geometry-dependent parameters of growth are found at each point. The computed resistance to heat diffusion $R_T$ is shown here. (c) Dwell times are calculated and written to a stream file, taking into account proximity effects between points in the same layer. (d) SEM of the printed model imaged at 45\degree\ tilt. Scale bar is 1$\upmu$m.}
\label{fig:algorithm}
\end{figure*}

We have implemented the layer-by-layer growth model explained above in a three-step algorithm designed to generate beam scanning patterns for FEBID deposition of 3D arbitrary geometries. The algorithm creates deposition sequences directly from STL files that can be designed in any standard 3D CAD software, the same approach followed by standard 3D printers.  This approach, in combination with the model described above, is a significant improvement with respect to recent FEBID works \cite{fowlkesHighFidelity3DNanoprintingFocused2018,kellerPatternGenerationDirectwrite2018}, simplifying the design and enabling the fabrication of complex 3D nano-objects. In this section we give an outline of the algorithm and demonstrate its effectiveness and flexibility. 

The workflow of the algorithm is described using the example of a free-standing concave surface (Figure \ref{fig:algorithm}a-c), with an SEM image of the fabricated nanostructure given in Figure 3d.
Firstly, a geometry is defined with an STL file which is sliced using constant-z planes, defining the layers of the structure (Figure \ref{fig:algorithm}a). 
In all experiments performed here, a maximum layer height of $\Delta z_l \leq 6$nm is set, to guarantee that the approximation of thin layers remains valid. 
Additionally, an adaptive slicing procedure has been implemented, adjusting the slice thickness based on the local angle of a structure. In this way, low hanging features are more finely defined, and a low lateral displacement between layers is maintained. From each slice, a dense mesh of dwell points is created, with points separated by a lateral ``pitch'' distance. We set both the displacement between layers, and the pitch to be 3nm. This value has been determined from dedicated experiments (see Supporting Information S7), where single nanowires were found to be widely independent of point pitch, for pitch values below 4nm, in agreement with literature.\cite{winklerHighFidelity3DNanoprintingFocused2018}

In a second step, geometry-dependent factors affecting the growth are calculated, and included as growth parameters for each dwell point. 
Specifically, the previously discussed resistance-based model for beam-induced heating is implemented, as displayed in Figure \ref{fig:algorithm}b. Further corrections for second-order effects such as beam defocussing and gas flux anisotropy are also possible, but only applied to objects where the added complexity is essential for successful deposition (see Supporting Information S5).  We anticipate that this second step of the algorithm could be expanded further by including additional factors relevant in other conditions and geometries, such as diffusion \cite{mutungaImpactElectronBeamHeating2019} and the dependence of secondary electron emission with local geometry. \cite{fowlkesFundamentalElectronPrecursorSolidInteractions2010} These, together with the measured calibration parameters, would redefine the model for deposition, $s(r)$, at every dwell point.

Thirdly, we implement the per-layer dwell time solver. Inter-point distances are computed and, together with the growth model, result in the intra-layer proximity matrix $s_{ij}$. The optimization problem for finding the appropriate dwell times at each point (as per Equation \ref{eqn:matrixdz}) is then solved using a non-negative linear-least squares solver. The colormap of Figure \ref{fig:algorithm}c  represents the resulting dwell times at each position. 

Finally, the beam scanning pattern is generated out of the computed dwells. In each layer, the beam is set to make multiple passes, reversing its direction from layer to layer in a ``serpentine'' pattern\cite{sanz-hernandezFabricationScaffoldBased3D2018}, to avoid exceeding a set maximum dwell time (5ms) and improving the smoothness of the deposition. The generated pattern is deposited, resulting in the SEM image shown in Figure \ref{fig:algorithm}d.

\begin{figure*}[ht!]
\centering
\includegraphics[width = 0.8\linewidth, draft=\draft]{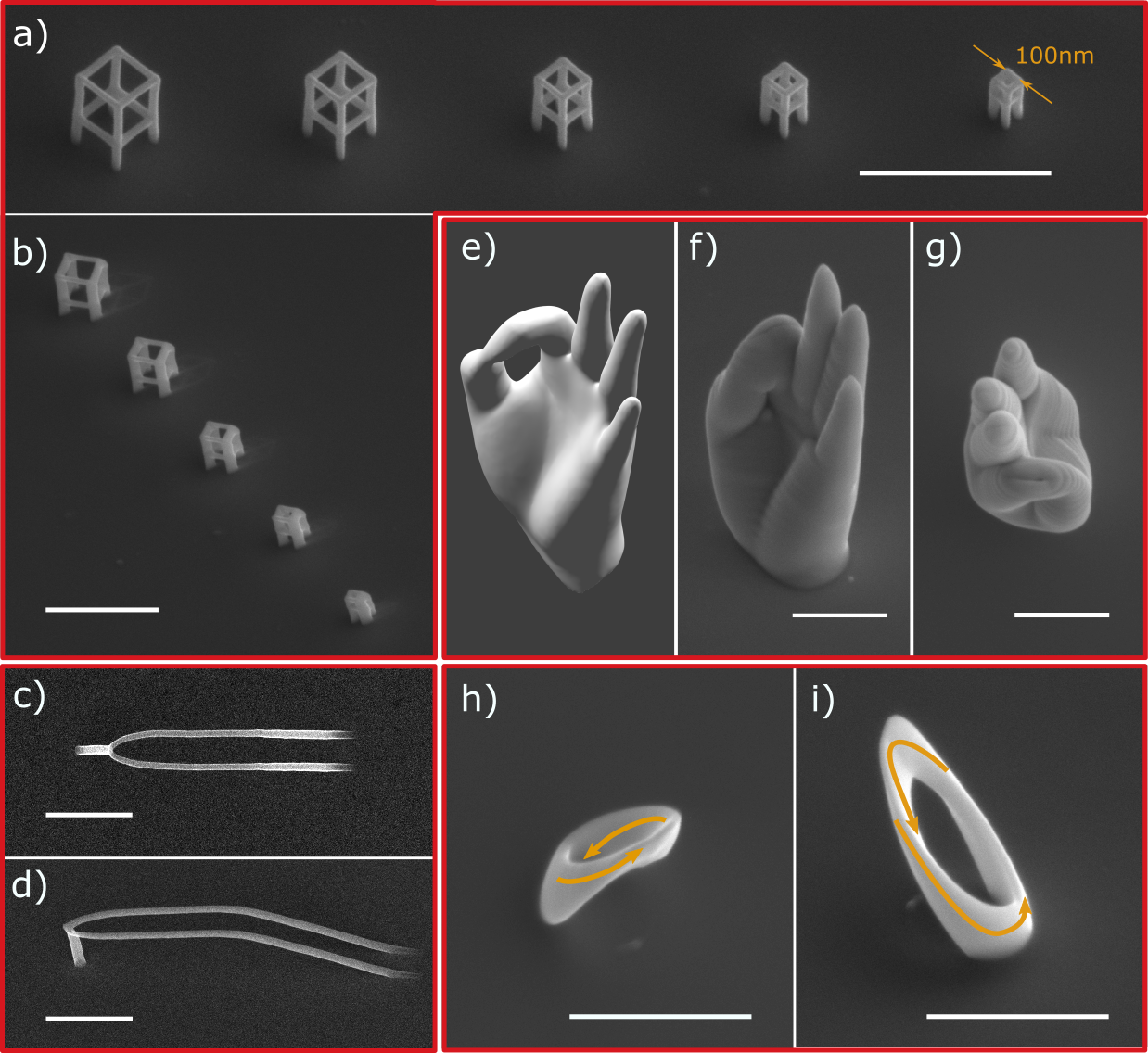}
\caption{SEM images under two orientations of a range of 3D geometries built using the three-stage 3D printing algorithm. (a-b) A series of cubes with side-lengths (left to right): 300, 250, 200, 150, 100nm at 45\degree\ tilt; (c-d) Smooth nanowires circuit imaged under 30\degree\ tilt. Vertical leg at the tip of the circuit is for added mechanical stability. (e-g) STL model, and side and top views of a nanoscale replica of a human hand.  (a-g) were built using a Pt-based precursor. (h-i) Cobalt M\"obius strip with arrows to help visualize the topology of the strip. The STL models of structures with further details on the viewing angles (S9), and a video of a M\"obius strip compiled from multiple images are provided in the Supporting Information. Scale bars are 1$\mu$m.}
\label{fig:structures}
\end{figure*}

In what follows, we demonstrate the capabilities of the algorithm by patterning a range of  geometries using two different (Pt- and Co-based) precursors. 
We first show that the fabrication of networks of straight nanowires can be reproduced, similar to the existing FEBID algorithms,\cite{fowlkesHighFidelity3DNanoprintingFocused2018,kellerPatternGenerationDirectwrite2018} by building a sequence of nanocubes of different sizes with half-pitch down to 50nm (see Figure \ref{fig:structures}a-b and Supporting Information S8 for further details).
Secondly, expanding recent achievements in nanowire device fabrication \cite{sanz-hernandezFabricationDetectionOperation2017,sanz-hernandezFabricationScaffoldBased3D2018}, in Figure \ref{fig:structures}c-d we create a  free-standing looped nanowire circuit formed by segments at different angles. The high quality and smoothness of the connections achieved here show the  potential of our method to nano-prototype advanced 3D nanoelectronic devices. \cite{sanz-hernandezFabricationDetectionOperation2017}.

Finally, the main advantage of our approach is the capability to fabricate arbitrary three-dimensional architectures out of 3D CAD files. As an example, we have fabricated a nanoscale human hand replica (Figure \ref{fig:structures}e-g) and a M\"obius strip (Figure \ref{fig:structures}h-i). These example structures present a wide range of features at different scales which are accurately replicated in experiments. In particular, Figure \ref{fig:structures}h-i shows the potential of this approach where, as far as we are aware, the smallest realization of a magnetic M\"obius strip has been fabricated. See the video in Supporting Information for a full characterization of this structure at multiple angles, and comparison with the corresponding  model file. 

We note though, that the thickening of some features is observed for the built structure with respect to the model, e.g. the hand (Figure \ref{fig:structures}f) and the associated STL file (Figure \ref{fig:structures}e). This edge thickening effect is expected when creating a volumetric object via FEBID due to the penetration of the primary electron beam through the structure, and subsequent generation of backscattered electrons and type II secondary electrons. In order to correct for this additional precursor dissociation, a ``shape-dimension adjustment technique''\cite{parikhCorrectionsProximityEffects1979} as the one employed in electron beam lithography could be implemented.

Another point to remark is that the model has been primarily developed using the standard Pt precursor due to its fast growth rate (of the order of 100nms$^{-1}$), low dependence on refresh times, and a well-known mechanism for decomposition under focused electron beams\cite{wnukElectronBeamDeposition2011}. Figure \ref{fig:structures}h-i demonstrate the algorithm's robustness when applied to other precursors. The Co-based precursor used to build this structure has a significantly more complex chemical behavior, including autocatalytic effects \cite{bishopFundamentalAdvancesFocused2019, muthukumarSpontaneousDissociationCo22012}. We attribute the transferability of our algorithm to complex gases to the ability of the basic Gaussian deposit model with a constant vertical growth rate, with no temperature correction:
\begin{equation}
  h(t, r) = GR_0\exp\bigg(\frac{-r^2}{2\sigma^2}\bigg)t
\end{equation}
to capture the effective average of deposition and give satisfactory first-order results for most regimes. This is especially true for wide structures (as opposed to single-pixel nanowires) because scanning along their width introduces a natural refresh rate. Moreover, beam heating effects are greatly reduced due to the higher heat conductivity provided by a larger volume, leading to more consistent growth rates (see Supporting Information S4.2).

\section{Conclusion}
We present a framework that makes possible high-fidelity layer-by-layer growth of complex-shaped 3D nanostructures using focused electron beams. 
Based on a layer-by-layer growth implemented in combination with the FEBID continuum model and proximity effect corrections, we can effectively account for a variety of effects, including beam-induced heating, defocussing, and gas anisotropy.
These effects have been studied for the Pt precursor, where we demonstrate how a large number of fundamental parameters can be reduced to only three, which are capable of effectively modeling the 3D deposition process. 

The framework has been implemented computationally to generate beam scanning patterns from STL files created by any standard 3D CAD software.
Using this approach, a wide range of nanogeometries which were until now inaccessible has been fabricated. These include surfaces with curvature along arbitrary directions, a M\"obius strip and a replica of a human hand. We have successfully tested our platform in three SEM systems and using two precursors with different physical and chemical properties, thus demonstrating the robustness and applicability of the technique.

This work paves the way for the advanced nano-manufacturing of 3D objects in a wide range of nanotechnology areas, making a fundamental step towards the study of advanced effects and their future exploitation.

\section{Experimental}
Three different dual-beam microscope systems were used for FEBID experiments: Helios 600 at the Wolfson Electron Microscopy Suite of University of Cambridge, FEI Nova 200, and  Helios 660 NanoLab at the Kelvin Nanocharacterisation Centre of University of Glasgow. Helios 600 and Nova 200 were used for deposition of \PtC, while Helios 660 was used with \CoC \ precursor. 

All depositions were performed on p-doped Si substrates under 21pA, 30kV electron beams. The algorithm for computing dwell points and corresponding dwell times was implemented in MATLAB using the built-in Optimization Toolbox. Computation times for all structures were in the order of $\sim10$s on most modern desktop computers, with the exception of the hand (Figure \ref{fig:structures}e-g) which took several minutes.  
The algorithm was in some cases supplemented with stage tilting to build features at various angles. In particular, this was exploited for the structures on Figure \ref{fig:structures}a-b(c-d) by building at 45\degree(30\degree) tilt in order to realize nanowires parallel to the substrate.
The fabrication times ranged from 2s for the shortest pillar in Figure \ref{fig:growth}b, 1min for the largest cube (Figure \ref{fig:structures}a-b), and up to 65min and 75min for the M\"obius and hand respectively (Figure \ref{fig:structures}e-g). For more information about the access to the software used in this paper, please contact the corresponding authors.

\pagebreak

\begin{acknowledgement}
This work was supported by the EPSRC Cambridge NanoDTC EP/L015978/1, an EPSRC Early Career Fellowship EP/M008517/1, and the Winton Program for the Physics of Suistanability. L. Skoric acknowledges support from St Johns College of the University of Cambridge. C.Donnelly  was supported by the Leverhulme Trust (ECF-2018-016), the Isaac Newton Trust (18-08) and the L'Or\'eal-UNESCO UK and Ireland Fellowship For Women In Science. Sara Merino-Aceituno was supported by the Vienna Research Groups grant number VRG17-014 by the Vienna Science and Technology Fund. Dédalo Sanz-Hernández acknowledges a Girton College Pfeiffer Scholarship.

The authors thank the staff of the Wolfson Electron Microscopy Suite of University of Cambridge and the Kelvin Nanocharacterisation Centre of University of Glasgow for technical support. We thank Jason D. Fowlkes, A. Hierro-Rodriguez and A. Welbourne for helpful discussions.

\end{acknowledgement}

\begin{suppinfo}
\begin{itemize}
	\item Experimental and technical details (PDF)
	\item Video showing the M\"obius strip from different angles compared to the corresponding 3D model (AVI)
\end{itemize}

\end{suppinfo}


\pagebreak
\bibliography{3dprinting}

\end{document}